\documentclass[aastex,12pt]{article}
\usepackage[]{aasms4}

\lefthead{M\"oller, Natarajan, Kneib \& Blain}
\righthead{Lensing by groups of galaxies}

\def\gs{\mathrel{\raise0.35ex\hbox{$\scriptstyle >$}\kern-0.6em 
\lower0.40ex\hbox{{$\scriptstyle \sim$}}}}
\def\ls{\mathrel{\raise0.35ex\hbox{$\scriptstyle <$}\kern-0.6em 
\lower0.40ex\hbox{{$\scriptstyle \sim$}}}}
\def\ltorder{
\mathrel{\raise.3ex\hbox{$<$}\mkern-14mu\lower0.6ex\hbox{$\sim$}}
}
\def\gtorder{
\mathrel{\raise.3ex\hbox{$>$}\mkern-14mu\lower0.6ex\hbox{$\sim$}}
}

\input psfig
\begin{document}

\title{Probing the mass distribution in groups of galaxies using
 gravitational lensing}
\author{Ole M\"oller$^{1,2}$, Priyamvada Natarajan$^{3}$, Jean-Paul Kneib$^4$ 
\& A.\,W. Blain$^5$}

\affil{1 Cavendish Laboratory, Madingley Road, Cambridge, CB3 OHE, UK }
\affil{2 Kapteyn Institute, PO Box 800, 9700 AV Groningen, The Netherlands}
\affil{3 Department of Astronomy, Yale University, 265 Whitney Avenue, 
New Haven, CT 06511, USA}
\affil{4 Observatoire Midi-Pyr\'en\'ees, UMR5572, 14 Avenue Edouard Belin, 
F-3100, Toulouse, France}
\affil{5 Institute of Astronomy, Madingley Road, Cambridge CB3 0HA, UK}

\begin{abstract}
In this paper, we study gravitational lensing by groups of
galaxies. Since groups are abundant and therefore have a large
covering fraction on the sky, lensing by groups is likely to be very
important observationally. Besides, it has recently become clear that
many lens models for strong lensing by individual galaxies require
external shear to reproduce the observed image geometries; in many
cases a nearby group is detected that could provide this shear. In
this work, we study the expected lensing behavior of galaxy groups in
both the weak and strong lensing regime. We examine the shear and
magnification produced by a group and its dependence on the detailed
mass distribution within the group. We find that the peak value of the
weak lensing shear signal is of the order of 3 per cent and varies by
a factor of about 2 for different mass distributions. These variations
are large enough to be detectable in the Sloan Digital Sky Survey
(SDSS). In the strong lensing regime we find that the image geometries
and typical magnifications are sensitive to the group properties and
that groups can easily provide enough external shear to produce
quadruple images. We investigate the statistical properties of lensing
galaxies that are near or part of a group and find that statistical
lens properties, like the distribution of time delays, are affected
measurably by the presence of the group which can therefore introduce
an additional systematic error in the measurement of the Hubble
constant from such systems. We conclude that both the detection of
weak lensing by groups and accurate observations of strong galaxy lens
systems near groups could provide important information on the total
mass and matter distribution within galaxy groups.
\end{abstract}

\section{Introduction}

Gravitational lensing by isolated galaxies and clusters of galaxies
has been used extensively both as a cosmological tool (Bartelmann 
et al. 1998; Wambsganss, Cen \& Ostriker 1998) and as a method to map 
detailed mass distributions (Mellier 1999). In the weak lensing regime,
reconstruction methods have provided moderate-resolution shear maps
(Hoekstra et~al. 1998; Clowe et~al. 2000; Hoekstra, Franx \& Kuijken 2000). 
In analyses that combine both weak and strong lensing data for clusters
it is found that further constraints can be obtained on the clumpiness
of the dark matter distributions on smaller scales within clusters
(Natarajan \& Kneib 1997; Geiger \& Schneider 1999).  This is done
exploiting the fact that image multiplicities and geometry are well
understood in the strong lensing regime for the HST cluster lenses,
and combining with ground-based weak shear data, mass models for the
critical regions of both clusters and the galaxies therein can be
constructed (Natarajan et~al. 1998).

Within the context of the current paradigm for structure formation --
gravitational instability in a cold dark matter dominated Universe
leading to mass hierarchies -- clusters and isolated galaxies are at
extreme ends of the mass spectrum of collapsed structures. Clusters
are the most massive, virialized objects, have the highest density
contrast and are rare, whereas galaxies are less massive and more
abundant. In this scenario, groups of galaxies, which lie in the
intermediate mass range between galaxy clusters and individual
galaxies, are the most common gravitationally bound entities at the
present epoch (Ramella, Pisani \& Geller 1997). Galaxy groups contain
between 3 and 30 galaxies and they trace the large-scale structure of
the Universe (Ramella et~al. 1999). The abundance of compact groups
was estimated by Barton et al. (1996) to be quite high, $1.4 \times
10^{-4}\,h^{-3}\,Mpc^{-3}$.  It is likely that they contribute
significantly to the mass density of the Universe,
$\Omega_{\mathrm{M}}$. Probing the mass of groups and the mass
distribution within groups is likely to be crucial to understanding
the evolution of both dark and baryonic matter.

In this paper we present numerical investigations of both the weak and
strong lensing signal expected from groups of galaxies. In
\S\,\ref{groups_props}, a brief overview of galaxy groups is
presented, concentrating on the properties of the observed compact
Hickson groups (Hickson 1982), followed by a
description of the properties of the simulated groups. In
\S\,\ref{groups_lensprops} the lensing properties of the mass models
is outlined along-with a brief description of the numerical analysis
techniques. \S\,\ref{groups_weak} focuses on the expected weak lensing
signal from groups of galaxies and its dependence on group properties.
In \S\,\ref{stronglens}, the effects of a group in the vicinity of a
strong galaxy lens are studied. The paper concludes with a summary of
the results and suggestions for a possible strategy for the
constraining the detailed mass distribution within galaxy groups in
\S\,\ref{groups_discuss}.

\section{Properties of Galaxy groups}
\label{groups_props}

The details of the distribution of dark matter in galaxy groups is
poorly understood at present. It is not known if there exists a common
dark matter potential well for the group as a whole or if the dark
matter content of the group is simply dominated by the individual
halos of the member galaxies. X-ray surface brightness profiles seem
to point towards the existence of a common group halo for compact
groups (Ponman et al. 1995; Helsdon et al. 2001), as extended and
diffuse X-ray emission is detected from the group as a whole, although
it is often observed to be centered around the optically brightest
galaxy in the group (Mahdavi et~al. 2000).

The precise morphology of the dark matter distribution in groups can
provide important theoretical constraints on their formation and
future evolution.  Gravitational lensing offers an elegant method to
map the mass content of the group and a first observational detection
has been reported by (Hoekstra et al. 2001).  For groups 50 selected
from the Canadian Network for Observational Cosmology Galaxy Redshift
Survey (CNOC2) at $z = 0.12 - 0.55$, they report a typical mass-to
light ratio in solar units in the B-band of 191$\pm$83 h, lower than
what is found for clusters indicating the presence of dark matter in
groups.  Since the lensing effects of individual galaxies are
detectable at a significant level (commonly referred to as
galaxy-galaxy lensing, see Brainerd, Blandford \& Smail 1996), we expect
an unambiguous signal to be obtained for groups. The prospect of
combining lensing data with X-ray data to probe the mass distribution
in galaxy groups is promising, specially in the light of current
high-resolution imaging X-ray satellites (Markevitch et~al. 1999;
Ettori \& Fabian 2000).

Detected galaxy groups have been found in two primary configurations:
compact and loose. The identification of groups and the establishment
of membership has proven to be difficult due to the ambiguity arising
from chance superpositions on the sky (Hickson et al. 1992; de
Carvalho \& Djorgovski 1995; Barton et al. 1996; Zabludoff \& Mulchaey
1998, 2000). Targeted spectroscopic surveys for groups are needed in
order to document and define the galaxies that constitute a group
(Humason et al. 1956; Barton et al. 1996; Ramella et~al. 1997).  Studies of the
morphological composition of compact groups (Williams \& Rood 1987;
Hickson et al. 1988) indicate that the fraction of late-type spiral
galaxies in these environments is significantly lower than in the
field. Additionally, there is evidence that the faint end of the
luminosity function of compact groups is depleted and the bright end
may be brightened (Barton et al. 1996).  Even when redshifts are known
for the group members, it is still sometimes unclear whether or not
they constitute a gravitationally  bound structure. From the velocity
histogram of the members while it often appears that the groups are
bound, they appear to be unvirialized (Zabludoff 2001). Extended X-ray
emission detected from hot plasma that is confined in the
gravitational well provides conclusive evidence for a group being
gravitationally bound (Mahdavi et~al. 2000; Helsdon \& Ponman 2000).

Compact groups have been cataloged by (Hickson 1982) and are easily
identified on the sky due to the high projected over-density of member
galaxies, but might not be truly representative of groups as a
whole. There has been some recent evidence from the studies of a large
sample of loose groups (Helsdon \& Ponman 2000) that the correlations
observed between their properties, such as X-ray luminosity, velocity
dispersion and temperature, differ from those of compact groups
studied by Mulchaey \& Zabludoff (1998).  The mass function of nearby
galaxy clusters follows a Press--Schechter form (Press \& Schechter
1974), and recent studies have shown that the mass function of loose
groups follows a similar distribution (Girardi \& Giuricin 2000),
suggesting a continuity of clustering properties from groups to rich
clusters. However, the detailed mass distribution in galaxy groups has
not been investigated so far. In particular, it is still unclear
whether the majority of groups possess a massive group dark matter
halo, as suggested by the observed X-ray emission of some
groups. Gravitational lensing allows us to precisely address this
fundamental issue. In this work, we study in detail the lensing
properties of the higher-redshift analogs of the Hickson compact
groups.

\subsection{Analytic forms for the potential}

The fiducial model studied here is a four-member compact group. The
total projected surface mass density at position $\vec{r}$ is simply
the sum of the surface mass densities of the individual galaxies,
$\Sigma_{\mathrm{i}}$, plus a larger-scale component
$\Sigma_{\mathrm{h}}$, that defines a larger scale group halo
encompassing all the individual galaxies:
\begin{equation}
\label{eq_summass}
\Sigma(\vec{r})={\Sigma_{\mathrm{h}}(\vec
  r)}+\sum_{i=1}^{4}\Sigma_{\mathrm{i}}(\vec{r}-\vec{r_{\mathrm{i}}}).
\end{equation}
Individual galaxies and the larger scale group halo are modeled as
scaled, self-similar PIEMDs (pseudo-isothermal elliptical mass
distributions; Kassiola \& Kovner 1993), parameterized by ellipticity
$\epsilon$, scale length $r_s$, truncation radius $r_t$ and central
density $\rho_0$. The projected surface density for such a model is :

\begin{equation}
\label{eq_mass}
\Sigma_i(\vec{r})= \Sigma_0 \frac{r_{\mathrm{s}}\,r_{\mathrm{t}}}
{r_{\mathrm{t}} - r_{\mathrm{s}}} \left 
(\frac{1}{\sqrt{r_0^2+k^2(x,y)}} -\frac{1}{\sqrt{r_{\mathrm{c}}^2+k^2(x,y)}}
\right),
\end{equation}
where $$k(x,y)=\sqrt{x^2/(1+\epsilon)^2+y^2/(1-\epsilon)^2}.$$

In the limit of a spherical halo, $\epsilon=0$, the projected mass
enclosed within radius $R$ is simply,
\begin{equation}
M(R)={\frac {M_{\mathrm{tot}}} {(r_{\mathrm{t}}-r_{\mathrm{s}})Rc^2}}\,
\left [\sqrt{r_{\mathrm{s}}^2+r^2}-\sqrt{r_{\mathrm{t}}^2+r^2}+
(r_{\mathrm{t}}-r_{\mathrm{s}})\right].
\end{equation}
where $M_{tot}=2\pi\Sigma_0 r_{\mathrm{s}} r_{\mathrm{t}}$ is the 
total mass, which is finite.

These self-similar PIEMDs provide a reasonable, realistic model of the
mass distribution in the both the large scale smooth component as well
as the mass associated with early type galaxies and has been used
previously successfully to model the mass profile of individual
galaxies in clusters by Natarajan \& Kneib (1996,1997).

\subsection{Properties of simulated groups}
\label{properties}

Each simulated group is defined by its redshift, $z_{\mathrm{l}}$, the
masses of the constituent galaxies, $M_{\mathrm{i}}$, their positions,
$\vec{r_{\mathrm{i}}}$, scale lengths, $r_{\mathrm{g}}$,
ellipticities, $\epsilon_{\mathrm{i}}$ and inclination angle measured
with respect to the x-axis, $\phi_{\mathrm{i}}$. The group halo is
assumed to be a spherical, pseudo-isothermal component with mass
$M_{\mathrm{h}}$ centered on the mass weighted mean position of the
individual galaxies. Real halos are likely to be elliptical in general
(as opposed to spherical), but since we will focus here on the average
lensing signal of a large sample of groups, this will not have a
significant effect on our results. Most defining parameters that
characterize the group properties are determined by drawing randomly,
from a probability distribution using a standard algorithm described
in Press et al. (1988).  Characteristic values for the drawn
parameters for the simulated groups are tabulated below.
\begin{table}
\begin{center}
\begin{tabular}{*{3}{l}}
\hline Parameter & Symbol & Value$^*$ \\ \hline 
Hubble parameter & $h$ & 0.5 \\
Matter density & $\Omega_{\mathrm{M}}$ & 0.3 \\
Cosmological constant & $\Omega_{\mathrm{\Lambda}}$ & 0.7 \\
Lens redshift & $z_{\mathrm{l}}$ & $0.3$ \\
Source redshift & $z_{\mathrm{s}}$ & $1.0$ \\
Number of group members & $N_{\mathrm{gal}}$ & 4\\
Group scale length & $r_{\mathrm{N}}$ & 15\,kpc-40\,kpc\\
Gaussian mean galaxy scale length & $\widetilde{r_{\mathrm{g}}}$ & 0.2\,kpc\\
Variance of galaxy scale length & $\sigma_{\mathrm{rg}}$ & 0.07\,kpc\\
Variance of galaxy ellipticities & $\sigma_{\epsilon}$ & 0.2\\
Mean galaxy mass & $\widetilde{M_{\mathrm{g}}}$ & $10^{12}\,\mathrm{M}_{\odot}$\\
Variance of galaxy mass & $\sigma_{\mathrm{M}}$ & $5\times10^{11}\,\mathrm{M}_{\odot}$\\
Gaussian mean halo scale length & $\widetilde{r_{\mathrm{h}}}$ & 15\,kpc\\
Variance in halo scale length & $\sigma_{\mathrm{rh}}$ & 3\,kpc\\
Cut off radius & $R$ & 100\,kpc\\
Fraction of total mass in halo & $f$ & 0-1 \\ \hline
* unless otherwise stated \\ \hline
\end{tabular}
\caption{List of the parameters used in this paper. Both cosmological
  parameters and parameters that define the group properties are
  listed.}
\label{table_props}
\end{center} 
\end{table}

\subsubsection{Group redshift}

The simulated groups were chosen to lie at a redshift of
$z_{\mathrm{l}}=0.3$, which corresponds approximately to the most
effective lens configuration for background sources with a mean
redshift of $z\sim1$. Due to the difficulty of detecting and
identifying groups, most known compact groups are at a redshift
substantially less than $z=0.3$, as mentioned previously, we are
concentrating here on the higher redshift analog of currently detected
groups at low redshift. However, new surveys, like the Sloan Digital
Sky Survey (SDSS), should provide a large number of groups at
redshifts of order 0.3. In \S\ref{projections} we show qualitatively
how variations in the redshift of the group, and possible projection
effects due to the effect of two groups lying at different redshifts
along the same line of sight affect our results. Note that for the
cosmological model of choice: $\Omega_{M}=0.3$ and
$\Omega_{\Lambda}=0.7$ an angular separation a $1''$ corresponds to
$6.24\,\mathrm{kpc}$ at a redshift of $z=0.3$.

\subsubsection{Galaxy positions}

The individual galaxy positions within each group are randomly
generated using a number density profile,
\begin{equation}
N(\vec{r})=\frac{N_0}{\left(1+r^2/r_{\mathrm{N}}^2\right)^{\beta}},
\label{eq_number}
\end{equation}
where $r_{\mathrm{N}}$ is the assumed typical group scale length. 
We use a value of $\beta=3/2$, which corresponds to the modified 
Hubble-Reynolds law everywhere in this paper except in \S 4.2. 
The normalization $N_0$ is determined by requiring,
\begin{equation}
\int_0^{\infty} N(\vec{r})\,d^2\vec{r}=N_{\mathrm{gal}}.
\end{equation}
In \S\,\ref{scale} the effects of the choice of the distribution of
the galaxy scale lengths and the form of the number density profile on
the results are discussed.

\subsubsection{Galaxy mass profiles}

\begin{figure}
\begin{center}
\psfig{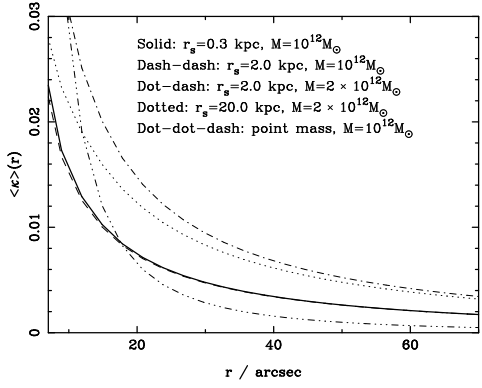}
\caption{The average surface mass density enclosed inside a circle of
radius $r$ for the mass profile given by eq.\, \ref{eq_mass}, compared to that
of a point mass. The mass M labeling the various style types is the total mass 
contained within a radius of 100\,kpc. Note that $\kappa\ll 1$ in all cases.}
\label{avkap}
\end{center}
\end{figure}

Group members are modeled by a PIEMD, given in eq.\, \ref{eq_mass}. A
suitable set of parameters for this profile are scale length
$r_{\mathrm{g}}$, total mass $M_{\mathrm{g}}$ enclosed within a radius
$R=100\,\mathrm{kpc}$, and an ellipticity $\epsilon$.  Fig.\,\ref{avkap}
shows the average enclosed surface mass density as a function of
radius for various choices of scale length and mass compared to the
point mass case.
Note that for small core radii the results do not depend strongly on
the choice of $r_{\mathrm{s}}$. To generate parameters for the
galaxies, we determine the scale lengths randomly from a Gaussian
distribution of mean $\widetilde{r_{\mathrm{g}}}$ and width
$\sigma_{\mathrm{rg}}$,
\begin{equation}
P_{\mathrm{rg}}(r_{\mathrm{g}})=\frac{1}{\sqrt{2\pi} 
\sigma_{\mathrm{rg}}}\,\exp\left[-{\frac{(r_{\mathrm{g}}-\widetilde{r_{\mathrm{g}}})^2}
{2\sigma_{\mathrm{rg}}}}\right].
\end{equation}
with the additional physical requirement that $0<r_{\mathrm{g}}$. The average
scale length is
$<\!r_{\mathrm{g}}\!>\,\approx\,\sigma_{\mathrm{rg}}/\sqrt{2}$. In a
similar fashion, we draw the ellipticities from a Gaussian
distribution with average $<\!\epsilon\!>=0$ and standard deviation
$\sigma_{\epsilon}=0.2$.  The mass $M_{\mathrm{g}}$ is similarly determined
randomly from a Gaussian distribution, with $0<M_{\mathrm{g}}$ so that the average mass
$<\!M_{\mathrm{g}}\!>\approx1/\sqrt{2}\sigma_{\mathrm{M}}$.  The
distribution of the inclination of the masses with respect to the
x-axis is assumed to be uniform. 

\subsubsection{Group halo}

The primary motivation for this study is to investigate the
possibility of determining the fractional mass of any common
intergalactic group halo that might be present in compact
groups. These group halos are also modeled using a PIEMD of the same
form as that for the member galaxies, eq.\,\ref{eq_mass}, centered on
the mean geometrical position of all galaxies. The parameters
describing the intergalactic halo are scale length, $r_{\mathrm{h}}$
and total mass, $M_{\mathrm{h}}$ within a cut-off radius $R$.  The
halo scale length is determined from a Gaussian distribution in the
same way as is done for the galaxies. The corresponding statistical
mean mass and standard deviation, $\widetilde{r_{\mathrm{h}}}$ and
$\sigma_{\mathrm{rh}}$ are respectively listed in
Table\,\ref{table_props}. The halo mass is determined by the masses of
the individual galaxies:
\begin{equation}
M_{\mathrm{h}}=f\times M{\mathrm{tot}}=\frac{f}{1-f}\times\sum_{i=0}^{4}M_{\mathrm{i}},
\end{equation}
where $M_{\mathrm{tot}}$ is the total mass in the group and $f$
denotes the total mass fraction in the halo, $0<f<1$.\\
Observationally, we are likely to have a better handle on the
masses of the individual galaxies than on the total mass of the group. 
We can then estimate the total mass of the group if we can obtain a value 
for $f$; this can in fact be done with weak lensing as is described below in
\S4. Alternatively, if the total group mass is determined then strong lensing 
could be used to constrain the mass of constituent members; this approach is
described in \S5.

\section{Lensing properties of Groups}
\label{groups_lensprops}

The total surface mass density $\Sigma$ induces a convergence $\kappa$
and shear $\gamma$ in the shapes of the background source population
located on a sheet at redshift $z_{\mathrm{s}}$.  We obtain
dimensionless forms for the surface mass density, potential and shear
in the usual way by defining the convergence
$\kappa (\vec{r})=\frac{\Sigma (\vec{r})}{\Sigma_{c}}$, 
scaled in units of the critical surface mass density
$\Sigma_{\mathrm{c}} = \frac{c^2D_{\mathrm{S}}}{4\pi GD_{\mathrm{L}}D_{\mathrm{LS}}}$,
where $D_{\mathrm{OS}}$, $D_{\mathrm{OL}}$ and $D_{\mathrm{LS}}$ are
the angular diameter distances from observer to source, from observer
to lens and from lens to source, respectively, as evaluated in a
smooth FRW Universe. The dimensionless form of the gravitational
potential can then be written as
\begin{equation}
\psi(\vec{r})=\frac{1}{\pi}\int\kappa(\vec{r}')\ln(|\vec{r}'-\vec{r}|)\,d^2r'.
\label{eq_potential}
\end{equation} 
For a pseudo isothermal sphere this potential can be written down
analytically:
\begin{equation}
\psi(\vec{r})=\frac{\Sigma_0r_0r_{\mathrm{c}}}{r_{\mathrm{c}}-r_0}\left[X-Y-r_0\ln{(r_0+X)}+r_{\mathrm{c}}\ln{(r_{\mathrm{c}}+Y)}\right].
\label{eq_piemd_potential}
\end{equation}
where $X=\sqrt{r_0^2+r^2}$ and $Y=\sqrt{r_c^2+r^2}$.
The components of the shear are given by,
\begin{eqnarray}
\gamma_1=\frac{1}{2}\left(\frac{\partial^2\psi}{\partial^2x}-
\frac{\partial^2\psi}{\partial^2y}\right)\\
\gamma_2=\frac{\partial^2\psi}{\partial x\partial y}=
\frac{\partial^2\psi}{\partial y\partial x}
\end{eqnarray}
and the magnification $\mu$ by,
\begin{equation}
\mu=\frac{1}{(1-\kappa)^2-\gamma^2}.
\end{equation}

We are concerned here with the measured shear produced by
gravitational lensing, the observable quantity is in fact the 
`reduced shear', which is a combination of $\kappa$ and $\gamma$,
\begin{equation}
\vec{g}=\frac{\vec{\gamma}}{1-\kappa},
\end{equation}
and is directly related to the induced ellipticity of a
circular background source (Bartelmann \& Schneider 2001). It is useful to quantify the 
tangential shear in terms of the components $\gamma_1$ and $\gamma_2$; for 
example to define an aperture mass (Kaiser 1995; Schneider et~al. 1998),
\begin{equation}
\gamma_{\mathrm{T}}=\gamma_1\sin{2\phi}+\gamma_2\cos{2\phi},
\end{equation}
where $\phi$ is the angle between $\vec{\gamma}$ and the x-axis of the
coordinate system.

\begin{figure}
\begin{center}
\psfig{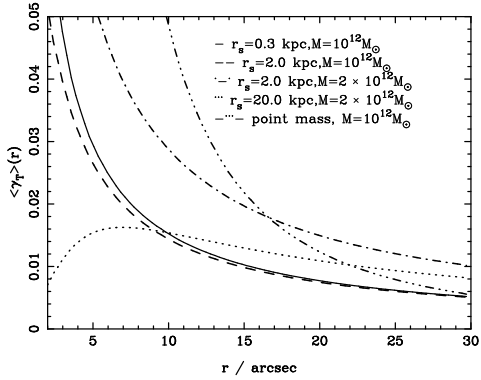}
\caption{The average tangential shear at radius r for the mass
profiles in Fig.\,\ref{avkap}.}
\label{avshear}
\end{center}
\end{figure}

Since our results are obtained using numerical simulations, we refrain
from presenting any further analytic formulae.  Fig.\,\ref{avshear}
shows the tangential shear produced by a spherical galaxy profile as a
function of radius (which is defined as the distance from the center 
of mass) for a range of scale-lengths and masses.
The magnitude of the shear at $r\approx10''$, of about 2 per cent, is
consistent with the findings of Brainerd et al. (1996) and Hoekstra et
al. (2001).  Note that the effect of a large core radius
$r_{\mathrm{s}}$ is to reduce the shear in the innermost regions
$r<r_{\mathrm{s}}$ below the value that is predicted at large
distances. Large core radii are not observed in galaxies (Cohn,
Kochanek, McLeod \& Keeton 2001), but extended group halos could in
principle possess large cores.

\subsection{Numerical methods}

The lens equation for groups is solved using the ray-tracing code
described in M{\"o}ller \& Blain (1998, 2001). With the exception of
\S\,\ref{projections} we use a single lens plane, as all group members
are assumed to have very similar redshifts. The deflection angle at
position vector $\vec{r}$ in the lens plane is then calculated
numerically from the expression of the surface mass density as given
in eqs.\,\ref{eq_summass} and \ref{eq_mass} using the formalism for
elliptical profiles developed by Schramm (1990). The adaptive grid
method as described in M{\"o}ller \& Blain (2001) is especially
suitable for the study of multiple lens systems such as groups of
galaxies, as it increases the achievable resolution around regions of
interest by a large factor.

\subsubsection{Weak lensing}

In order to compute weak lensing by groups of galaxies, we generate a
fine grid of $N=n_x\times n_y\sim 10^6$ pixels which are assigned
reduced shear values obtained from a numerical ray-tracing
simulation. From this fine grid, we determine the reduced shear profile
for different group models. The numerical error due to the simulations
is negligible.

\subsubsection{Strong lensing}

The magnification maps on the source and image plane are obtained
using the ray tracing of triangles as described in Schneider, Ehlers 
\& Falco (1992) and M{\"o}ller \& Blain (2001). The number of
images on a regular grid in the source plane are also obtained using the
same ray-tracing routines. For every point on this grid, we store the number 
of images, the image positions, magnifications and time delays for each individual 
image. This information is used to obtain the statistical properties presented 
in \S\ref{stronglens_stat}. The time delay $\Delta T$ between two images,
at $\vec{\theta_1}$ and $\vec{\theta_2}$, of a source at $\vec{\beta}$
is obtained using the equation,
\begin{equation}
\Delta T=C\times(1+z_{\mathrm{l}})\left[\frac{(\vec{\beta}-\vec{\theta_1})^2-(\vec{\beta}-\vec{\theta_2})^2}{2}+\Delta\psi\right].
\end{equation}
where $C=(D_{\mathrm{OS}}D_{\mathrm{OL}})/(cD_{\mathrm{LS}})$, $\Delta\psi=\psi(\vec{\theta}_2)-\psi(\vec{\theta}_1)$ and the potential $\psi$ itself is given by eq.\,\ref{eq_potential}.

\section{Weak Lensing by groups}
\label{groups_weak}

Weak gravitational lensing provides an extremely useful tool to map
mass distributions on large scales, ranging from a few hundred
kiloparsecs to a megaparsec. The shear of background galaxies around
clusters, which in the weak regime is at the 1\% level has been used
quite successfully to determine the cluster potential (Hoekstra
et~al. 1998; Fischer 1999; Clowe et~al. 2000); but, because of their
smaller mass, the signal from galaxy groups is expected to be much
lower. The mass contrast from groups is similar to or greater than
that from large scale structure, and so recent progress in sensitivity
and methods has made the detection of weak lensing signals by groups
feasible (Hoekstra et al. 2001).  An individual compact group occupies
a small area on the sky, therefore the essential limitation is due to
the small number of background galaxies that lie directly behind the
group, therefore to detect a signal several groups have to be stacked
akin to the case of galaxy-galaxy lensing (Brainerd, Blandford \&
Smail 1996) in order to increase the signal-to-noise ratio. The
distinguishable effects of the choice of different group mass profiles
on the resultant averaged, stacked shear map from a sample of about 100
random groups is studied in this section.

\subsection{Distinguishing group halo vs. individual halos}

\begin{figure}
\begin{center}
\psfig{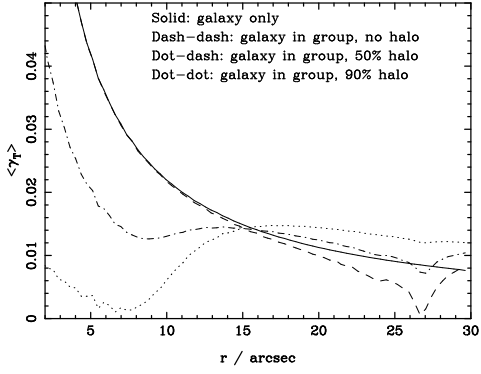}
\caption{The average tangential shear computed centered at a group member, 
for different values of the halo:galaxy mass ratio. The total group mass
is the same in all cases. The solid line shows the shear for an
isolated galaxy for comparison.}
\label{avmember}
\end{center}
\end{figure}

\begin{figure}
\begin{minipage}{140mm}
\hskip -5mm (a) \hskip 67mm (b)
\begin{center}
\vskip -5mm \hskip -10mm \psfig{file=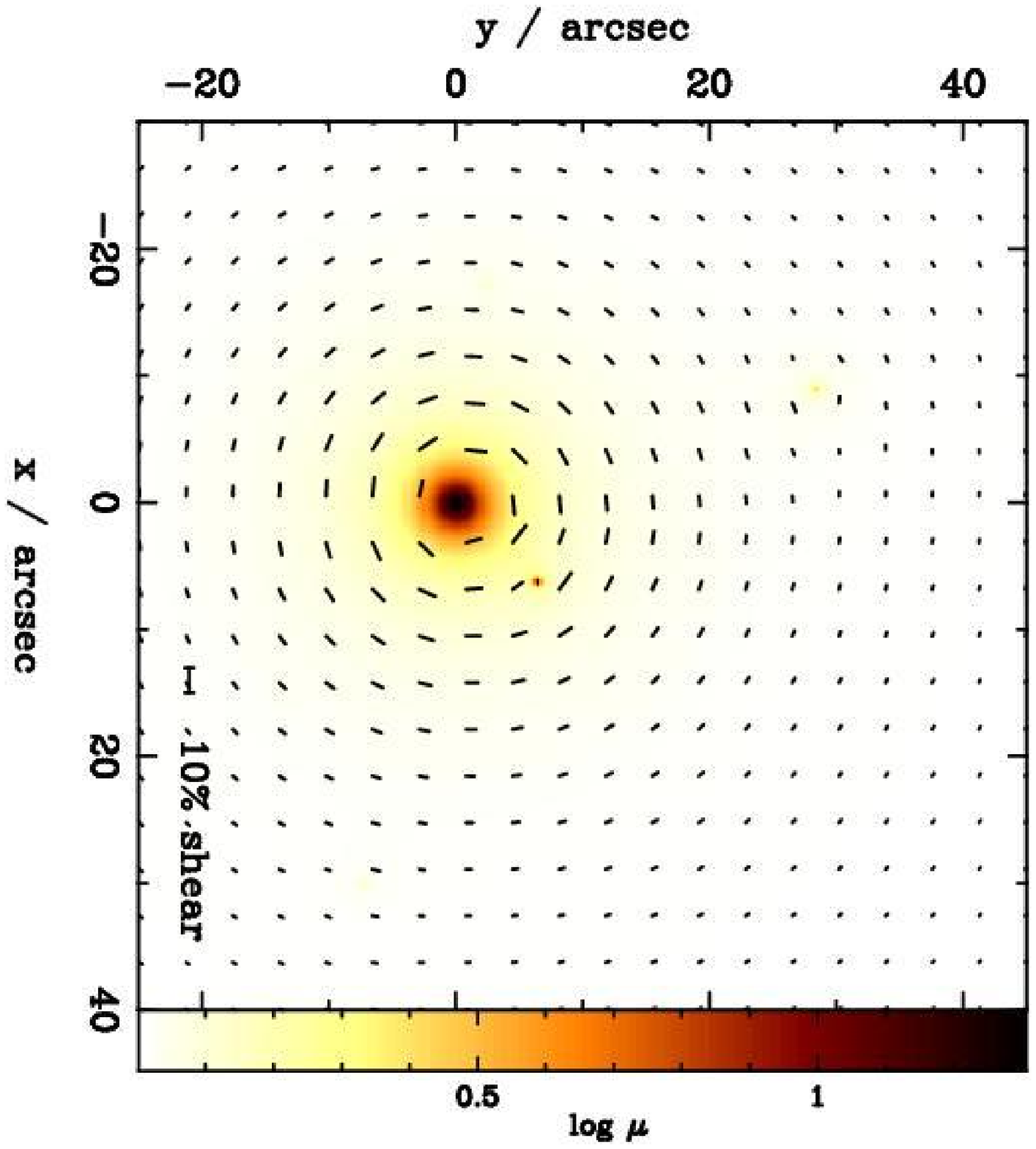,width=6.5cm,angle=-90} \hskip 0mm\psfig{file=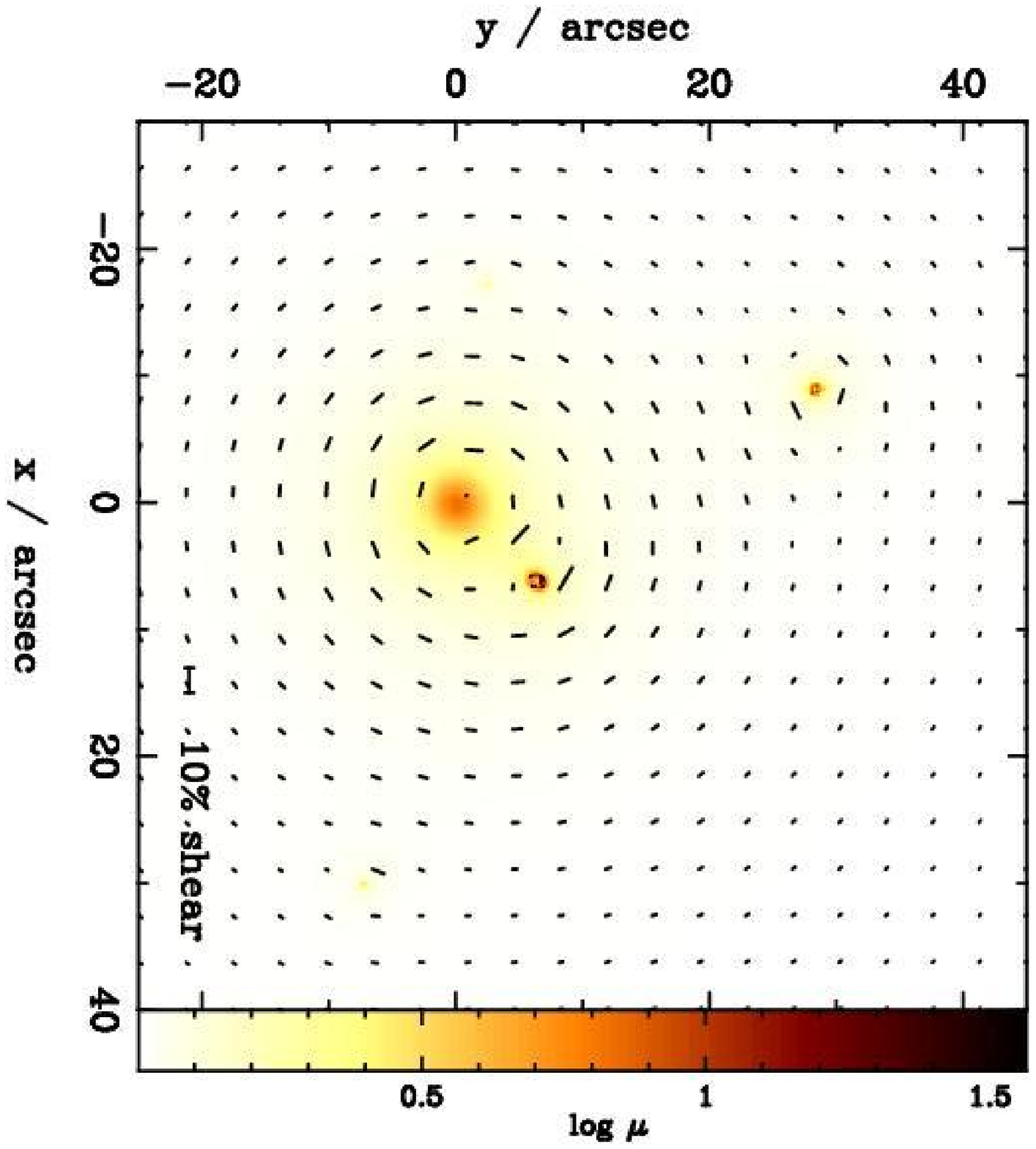,width=6.5cm,angle=-90}
\end{center}
\hskip -5mm (c) \hskip 67mm (d)
\begin{center}
\vskip -5mm \hskip -10mm  \psfig{file=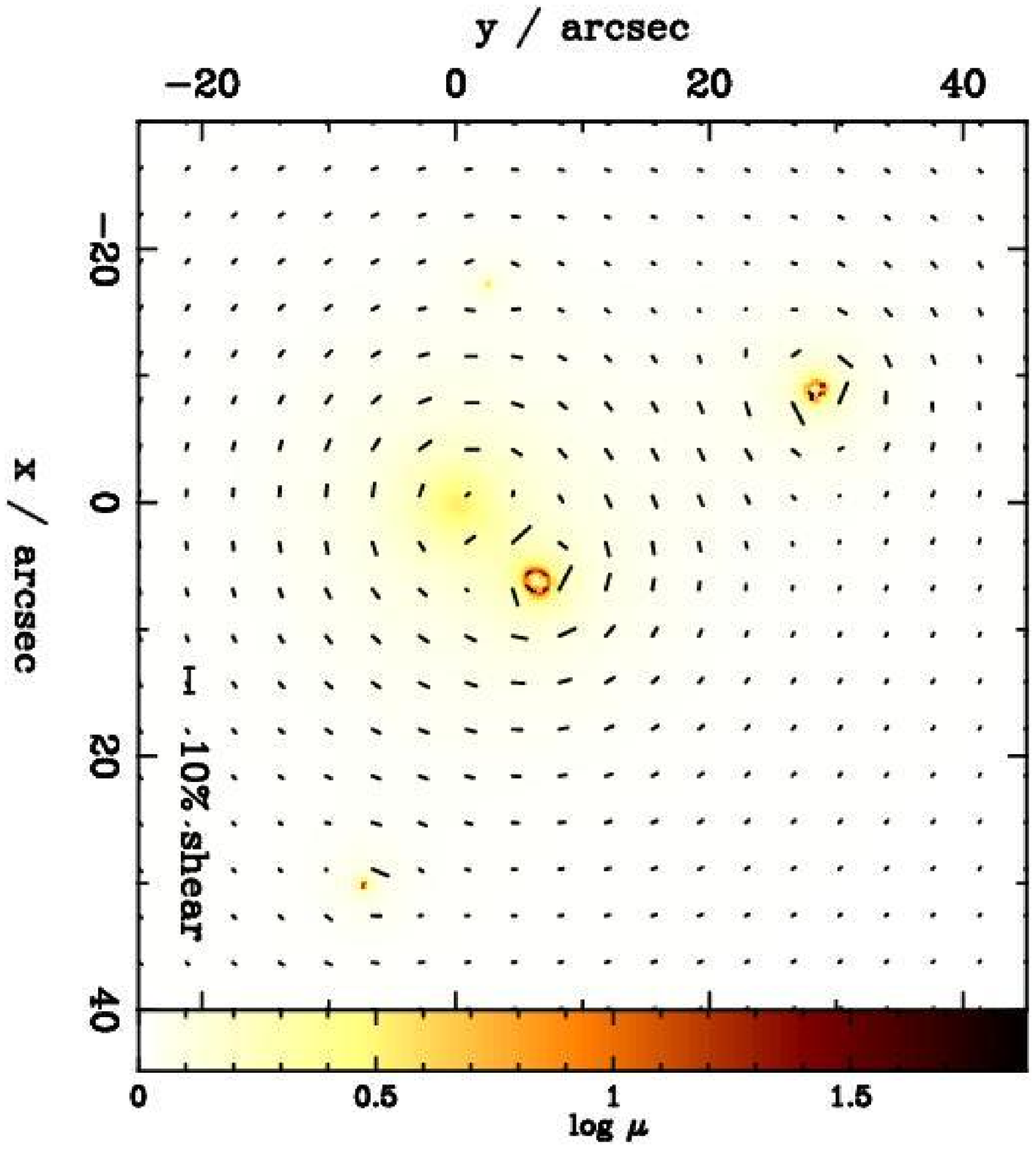,width=6.5cm,angle=-90}\hskip 0mm\psfig{file=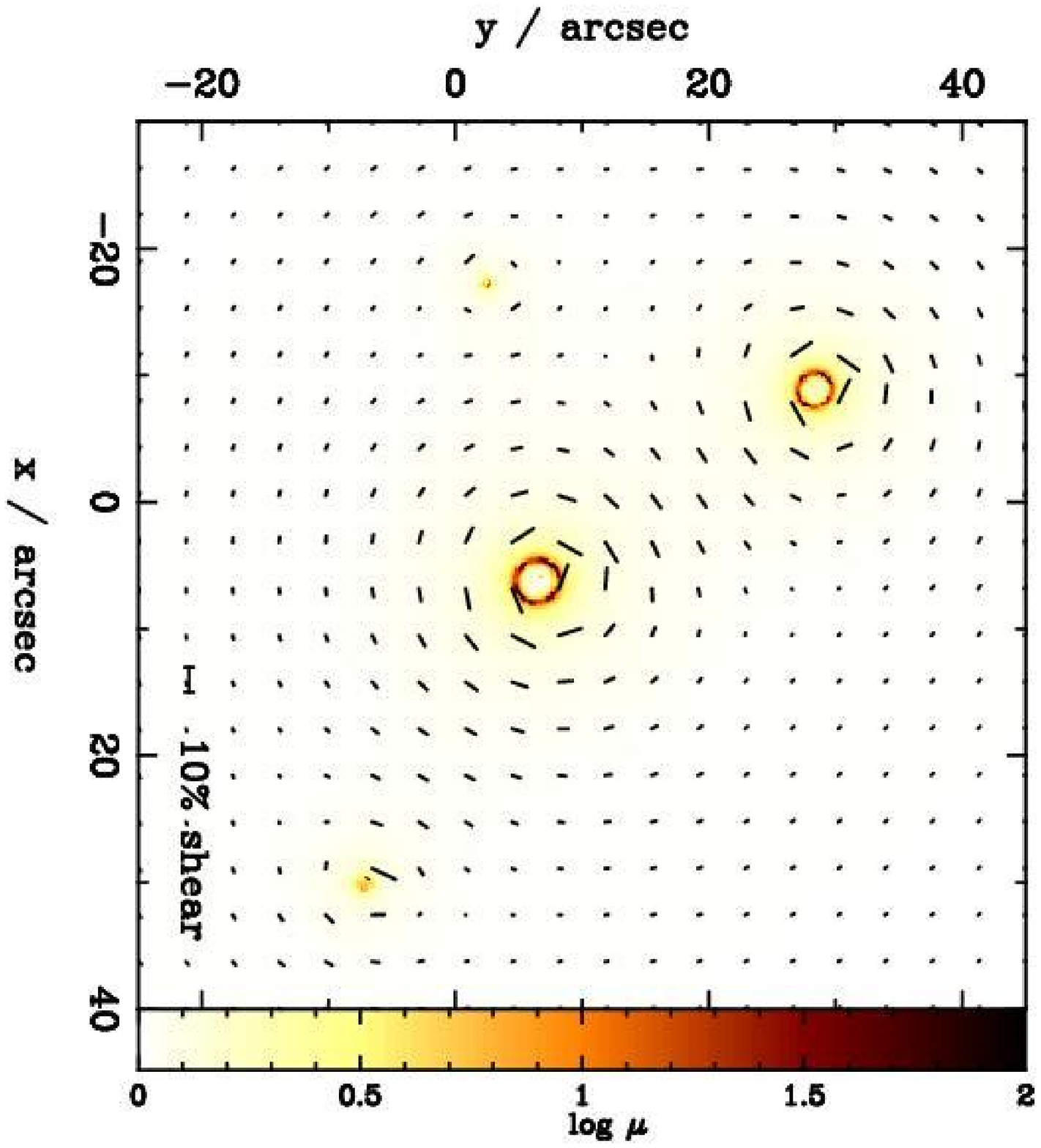,width=6.5cm,angle=-90}
\end{center}
\caption{Shear and magnification maps for a typical compact group with 4
member galaxies. The panels show the magnification on the image plane
as a grey-scale. The direction and relative magnitude of the shear is
indicated by the arrows in each panel. Panel (a) is for a group
corresponding to Model A in table \ref{table_models}, panel (b) for a group 
corresponding to Model B and panel (c) for a group corresponding to Model C. 
Panel (d) shows the shear and magnification for a group as in Model D, 
with no halo. In all panels, the source plane redshift is taken to be 
$z_{\mathrm{s}}=1.0$. }
\label{maps}
\end{minipage} 
\end{figure}

\begin{figure}
\begin{center}
\psfig{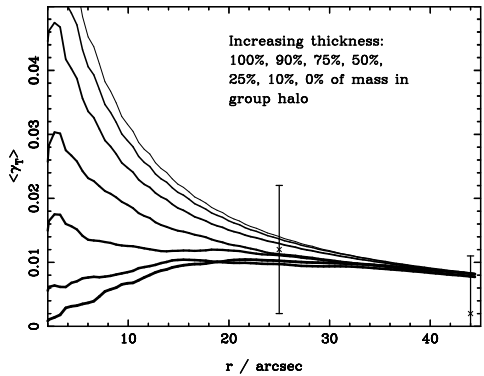}
\caption{The tangential shear averaged over 100 groups for different 
values of the halo mass: galaxy mass ratio. The total group mass is the 
same in all cases. The overplotted observational data point is from
Hoekstra et al. (2001) for the CNOC2 groups.}
\label{avhalo}
\end{center}
\end{figure}

In fig.\,\ref{avmember} the effect of varying the ratio of mass in the group
halo to that associated with individual group member galaxies is plotted 
for the detected shear signal centered around an individual member. 
Increase the fraction of mass attributed to the group halo leads to a 
lowering of the shear signal at small radii. The reduction of the shear in
the inner regions is primarily due to the relatively small mass contribution 
of the individual galaxies and is compensated by the increase in the external 
shear produced by the presence of the halo, introducing an asymmetry 
in the shear pattern. This is a generic effect, which is found in the lensing 
signal of all member galaxies, but its strength varies depending on the relative 
positions of the galaxies and halo. In Fig.\,\ref{maps}(a)-(d) we show shear
and magnification maps for a group varying the halo to galaxy mass
ratio keeping the total group mass fixed.
\begin{table*}
\begin{center}
\begin{tabular}{*{6}{l}}
\hline Parameter & Halo & Galaxy 1 & Galaxy 2 & Galaxy 3 & Galaxy 4\\ \hline 
x-position & 0'' & 30.0'' & -8.9'' & 6.2'' & -17.3''\\
y-position & 0'' & -7.2'' & 28.3'' & 6.4'' & 2.4''\\
M in 100\,kpc / $10^{11}M_{\odot}$ & & & & & \\
--Model A & $155$ & $2$ & $6$ & $8$ & $1.2$\\
--Model B & $120.7$ & $6$ & $18$ & $23.9$ & $3.6$\\
--Model C & $86.2$ & $10.1$ & $30.1$ & $40$ & $6$\\
--Model D & $0$ & $20.2$ & $60.1$ & $80$ & $12$\\
Redshift $z_{\mathrm{l}}$ & 0.3 & 0.3 & 0.3 & 0.3 & 0.3 \\ 
$r_{\mathrm{s}}$ / kpc & 15 & 0.3 & 0.5 & 0.7 & 0.1\\
\hline
\end{tabular}
\caption{Individual group models. The properties of the individual
  group models used in \S\,\ref{stronglens} and Fig\,\ref{maps}. }
\label{table_models}
\end{center}
\end{table*}
The galaxy substructure is clearly evident in all the shear and magnification 
maps. However, the figure also illustrates that this substructure is unlikely 
to be detectable for an individual group; the shear signal is too small.  In
order to make a positive detection, many groups will have to be
stacked. Fig\,\ref{avhalo} shows the sensitivity of the inner slope of
the shear profile for an average of 100 stacked groups with varying
group halo to galaxy mass ratio.

\begin{figure}
\begin{center}
\psfig{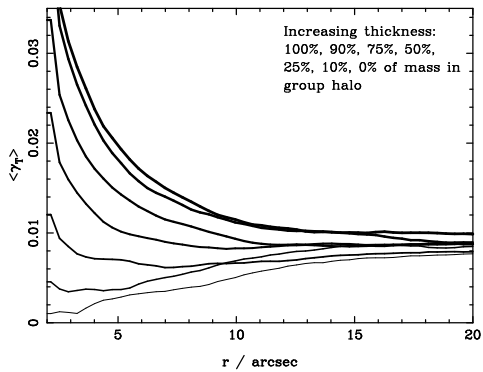}
\caption{The average tangential shear around the individual group members of 100 
simulated groups for different halo : galaxy mass ratios. The qualitative behavior is 
similar to that in Fig.\,\ref{avhalo}; the signal at small radii is strongly dependent 
on the relative masses of galaxy and group halos. Note that the difference between this plot 
and the previous one lies in the choice of center around which the shear field is averaged.}
\label{avind}
\end{center}
\end{figure}

The shear signal within 20'' varies by about 1 percent, which is once again
detectable when averaged over 100 groups. There is however some uncertainty, 
as we have do not have a priori knowledge of the profile slope and core 
radius of the intergalactic group halo. A possibility would be to use the 
X-ray profile, and assume that the same form describes the mass
distribution in the halo. Another difficulty arises from the fact that
it is necessary to determine the position of the `center' of the group
in order to stack the tangential shear signal coherently. If either this
determination is inaccurate or the intergalactic halo is off-center
from this position, then the measured shear profile will be flatter,
leading to a systematic underestimate in the halo mass fraction.  An
elegant solution to these problems is to add the average tangential
shear around each member galaxy. The positions of galaxies can be
determined accurately from their light distribution. Furthermore, the
slope of the mass profile of the individual galaxies is much better
constrained from the studies of individual lenses. Since the
constraints on the positions and profiles of the individual galaxies
are likely to be tighter, the average shear profile around each of the 
member galaxies can be related directly to the group halo to galaxy mass 
ratio. Determining the shape of the average shear profile around a set of 
galaxies can therefore provide a new, important and feasible way of determining 
the relative mass fraction in galaxies in different environments.
Fig.\,\ref{avind} shows the resulting shear profile around the member
galaxies averaged over 100 groups.

Qualitatively, the same effect is seen in both cases (independent of
choice of center), a strong correlation between the relative mass
distributions and the average value of the shear; however for the case
of massive halos, the average shear around group members is
significantly reduced at small radii. In the following sections the
average shear around the member galaxies, rather than the shear around
the ill-defined group center, will be considered.

\subsection{Dependence on the density profile of galaxies in the group}
\label{scale}

\begin{figure}
\begin{center}
\psfig{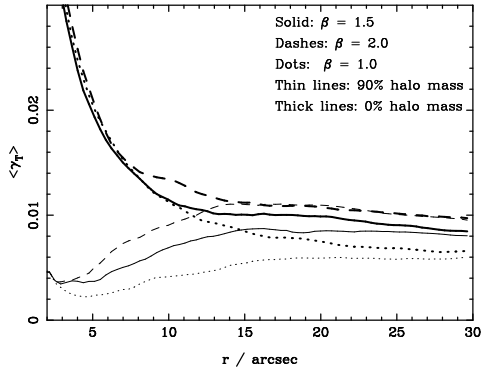}
\caption{The average tangential shear around the individual group
members of 100 simulated groups for different shapes of the galaxy
number density profile, eq.\,\ref{eq_number}. The groups are generated
randomly as described in section \ref{properties} and are all at
$z_{\mathrm{l}}=0.3$.}
\label{avdens}
\end{center}
\end{figure}

The galaxy profile given in eq.\,\ref{eq_mass} has been used
extensively and provides a good approximation to the true mass profile
of most galaxies. Furthermore, past studies have
shown that galaxies have a compact core radii, and as shown in
Fig.\,\ref{avshear}, the observed shear variations with galaxy core
radius and ellipticity is expected to be small. The effect of the
choice of the form of the galaxy number density profile on the measured 
shear is shown in Fig.\,\ref{avdens}.

The figure shows that for number density profiles steeper than the
modified Hubble-Reynolds law, the average tangential shear at radii of
about 30 arcseconds is increased relative to the inner average
shear. This is due to an increase in the average mass density both
inside the group and around member galaxies for more spatially compact
groups.

The analysis presented here mainly concerns the study of small compact
groups.  We also performed simulations for group scale lengths between
15 and 40\,kpc and found that the shear profile does not vary
significantly; this is as we expect given that we normalize the mass
at a relatively small radius of $100\,\mathrm{kpc}$.

Recently, there has been much discussion about the NFW (Navarro, Frenk \& 
White 1996) density profile -- which has been fit successfully in N-body
simulations to dark matter halos on a large range of scales from
small galaxies to rich clusters. The lensing properties of this profile have
been studied recently by (Wright \& Brainerd 2000). We explore this 
density profile for group members in our simulations. The NFW profile has 
the form,
\begin{equation}
\rho(r)=\frac{\rho_0}{r/r_{\mathrm{s}}(1+r/r_{\mathrm{s}})^2},
\end{equation}
where $r_{\mathrm{s}}$ is a characteristic scale length and $\rho_0$
is a central density. The total mass interior to radius $R$ for an
NFW profile is
\begin{equation}
M(r)=4\pi\rho_0 r_{\mathrm{s}}\left(\ln ({1+\frac{r}{r_{\mathrm{s}}}})-\frac{r}{r+r_{\mathrm{s}}}\right).
\end{equation}
Analytical expressions for the shear can be found in Wright \&
Brainerd (2000) and Trentham, M{\"oller \& Ramirez-Ruiz (2000). The
scale lengths for this galaxy model are assigned randomly in exactly
the same way as for the PIEMD as described in section
\ref{properties}. The total mass inside a radius of
$100\,\mathrm{kpc}$ is set to the same value as for the PIEMD. Since
the NFW profile is shallower inside $r_{\mathrm{s}}$ and steeper
outside that radius as compared with the PIEMD profile, the mass at
small radii is larger than that for the equivalent PIEMD. The shear 
signal is therefore expected to be larger at small radii. Fig.\,\ref{avnfw}
shows the tangential shear around a group in which all the mass
components are modeled as NFW profiles.

\begin{figure}
\begin{center}
\psfig{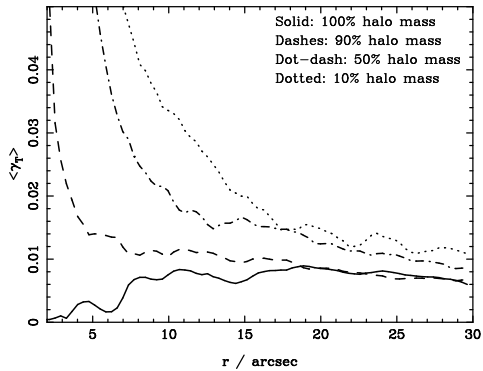}
\caption{The average tangential shear around the individual group members of 100
simulated groups for an NFW galaxy and halo profile. The group
properties are determined randomly as for Figs.\,2-5. All groups are
at $z_{\mathrm{l}}=0.3$ and the source redshift $z_{\mathrm{s}}=1.0$.}
\label{avnfw}
\end{center}
\end{figure}

Results are qualitatively similar to those obtained for the PIEMDs (in
Fig.\,\ref{avmember}).  However, the effect on the shear field of
shifting mass from the halo to the individual galaxies is much more
pronounced due to the larger mass contained at small radii in the NFW:
the central value of the shear is even more sensitive to the halo mass
to galaxy mass ratio, and the shear at small radii is generally
larger.

\subsection{Dependence on the redshift distribution of the group members}
\label{projections}

Uptil now all group members were assumed to lie at the same
redshift. For groups that have been selected from a spectroscopic
survey with accurate redshift determinations that will indeed be the
case. However, it is instructive to investigate any qualitative
differences that might arise in the shear profiles due to projection 
effects. Fig.\,\ref{avredshift} shows the average shear around member 
galaxies for three cases:
\begin{figure}
\begin{center}
\psfig{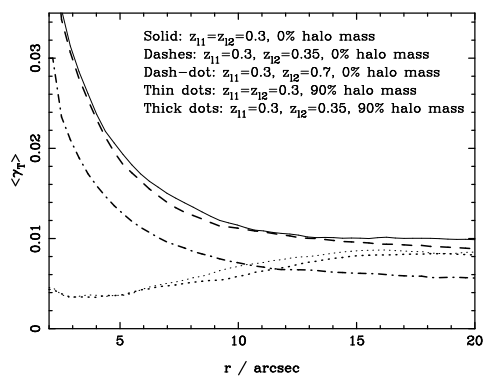}
\caption{The average tangential shear around group members of 100
simulated groups for different lens redshifts. Results for projected
groups in which the galaxies are at different redshifts are shown
along with the results for compact groups at a single redshift 
of $z_{\mathrm{l}}=0.3$.}
\label{avredshift}
\end{center}
\end{figure}
(i) in which all member galaxies are at the same redshift; (ii) where
half of the `member' galaxies are at a much higher redshift; and (iii)
in which the redshift difference is $\Delta z/z\sim0.2$. As expected,
a small redshift difference does not lead to significantly
different results, whereas a larger redshift difference, with only
part of the apparent group lying at an optimum lens redshift, leads to
a decrease in the shear signal by a factor of a few. Note that the
shape of the shear profiles is not affected by differences in redshift
and that the measured ratio of the tangential shear at small radius to
that at large radius is therefore still a good estimator of the
relative halo : galaxy mass ratio. However, since the overall shear
signal is reduced if galaxies are mistakenly assumed to be part of a
group, the total mass in the group is likely to be systematically 
underestimated.

\section{Strong lensing effects}
\label{stronglens}

In the previous section we calculated the expected magnitude of the
weak lensing signal due to galaxy groups. In the strong lensing
regime, there is clear observational evidence for the effect of the
group potential. Many of the known lens systems cannot be described
accurately by a single lens model and a significant external shear is
required in many cases (Kundic et~al.  1997a; Keeton, Kochanek \&
Seljak 1997; Kneib, Cohen \& Hjorth 2000). In fact, groups of galaxies are found near many of these
systems (Rusin et al. 2000). In general, the presence of a group in the vicinity
of a galaxy lens will lead to an external shear contribution to the
main lensing potential. The direction and magnitude of this shear will
depend strongly on the precise mass distribution within the group and
this will affect the image positions, magnifications, time delays and
image geometries. This will be important for modeling both the 
individual lens systems as well as for lens statistics (number of multiple
images produced).

\subsection{Individual lens systems}

A basic consequence of the presence of a group on the lensing behavior
of a nearby galaxy is to introduce some external shear. As shown for
example by Keeton, Kochanek \& Seljak (1997), the effect of this shear
is to introduce an effective asymmetry in the potential which affects
the image geometries and magnification ratios. In the following we
investigate the effect of the details of the mass distribution inside
the group which neighbors an individual lensing galaxy.

\subsubsection{Magnification and image geometry}

\begin{figure}
\begin{minipage}{140mm}
\hskip -5mm (a) \hskip 67mm (b)
\begin{center}
\vskip -5mm \hskip -10mm \psfig{file=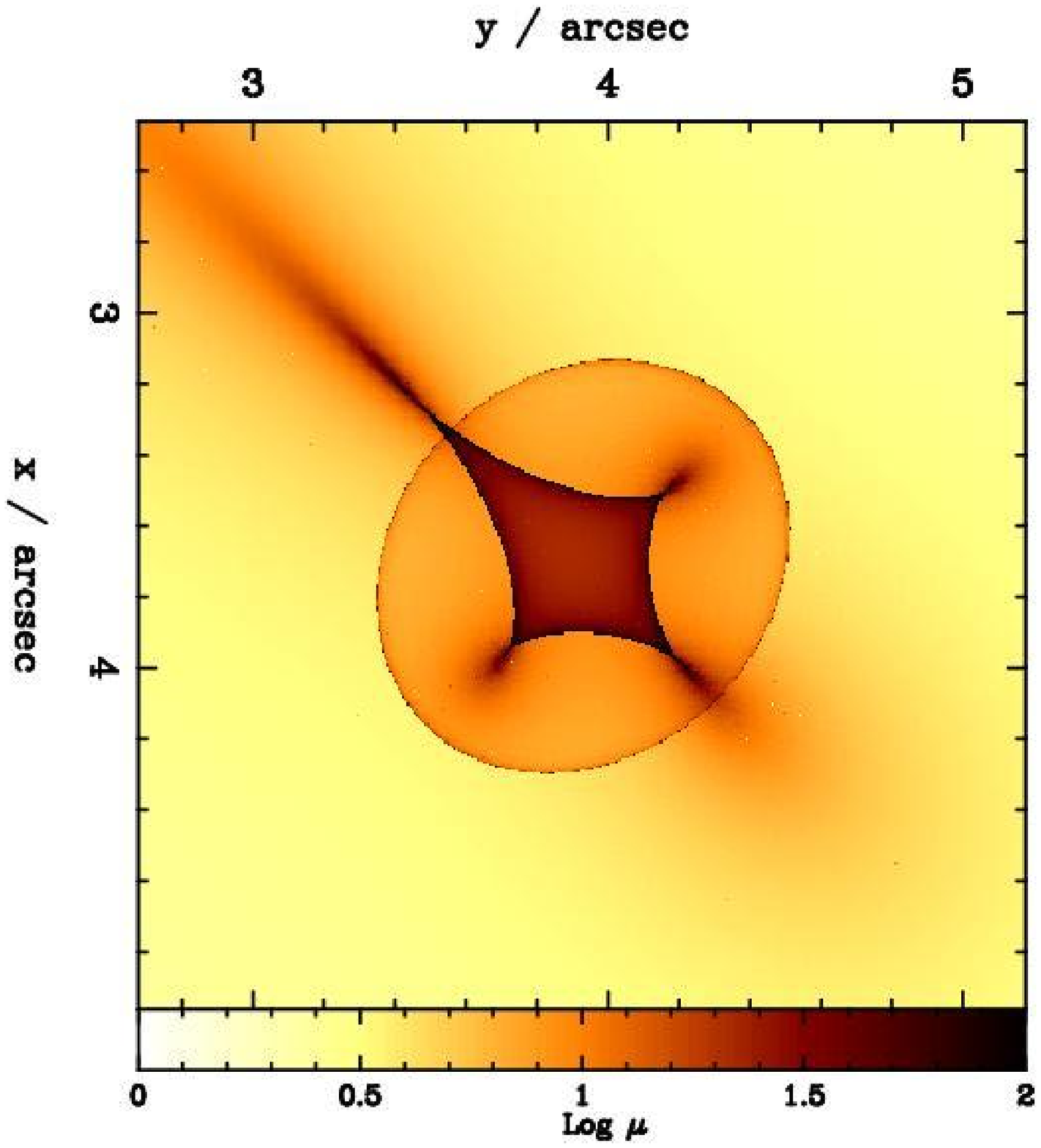,width=6.5cm,angle=-90} \hskip 0mm
\psfig{file=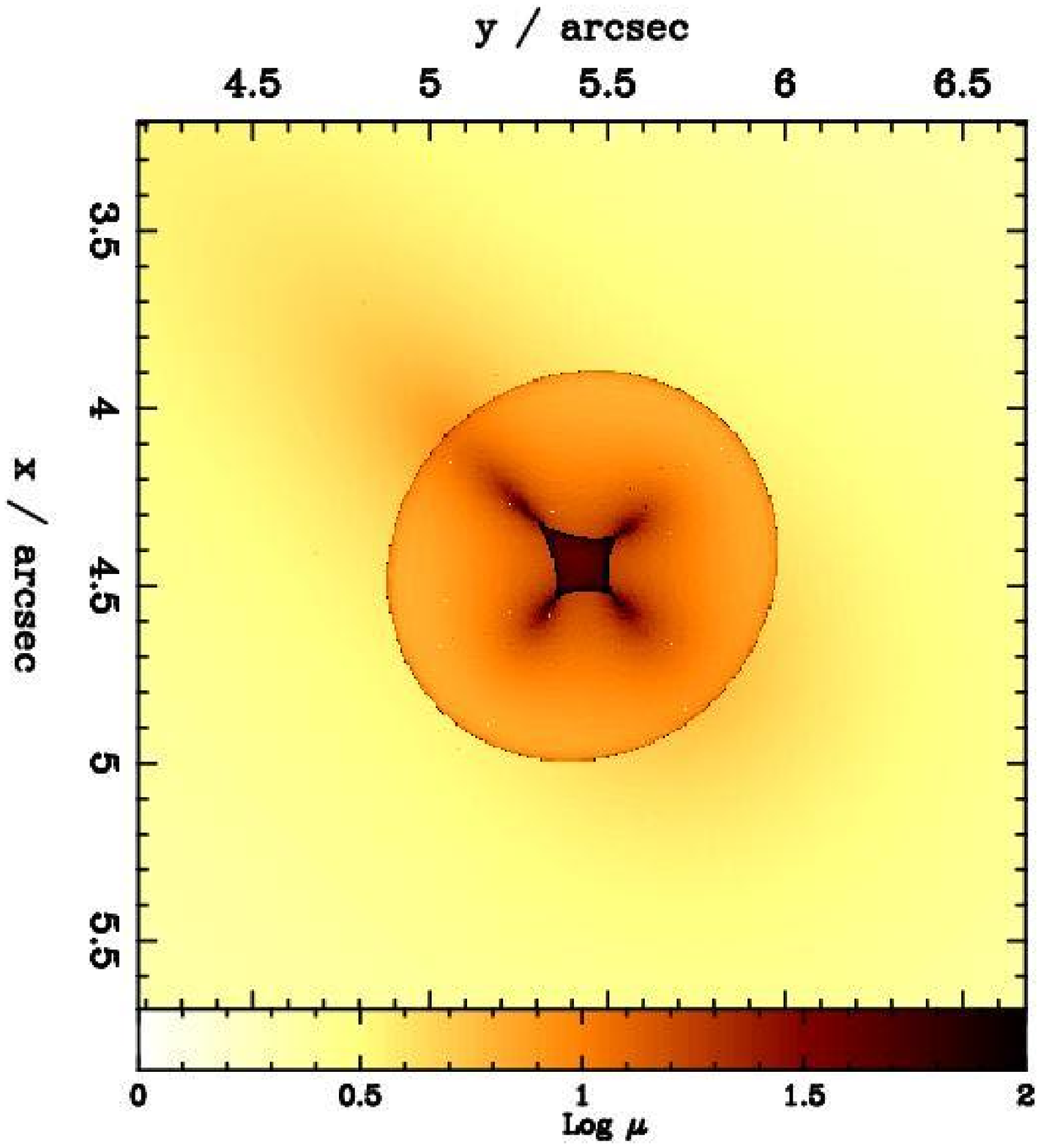,width=6.5cm,angle=-90}
\end{center}
\caption{Magnification maps on the source plane for a typical compact
group with four member galaxies. The panels are centered on the
position of the main lensing galaxy, which has the same properties as
galaxy 3 of Model C (tab.\ref{table_models}) in both panels. The
properties of the group members vary; in the left panel, the group
properties are those of Model A; in the right panel, the group
properties are those for Model C; Note that the properties of the main
lensing galaxy are the same in both panels and the differences in the
maps arise solely from the variation of the mass distribution in the
surrounding group.}
\label{mapsstrong}
\end{minipage} 
\end{figure}

Many of the expected strong lensing properties of particular lens
profiles can be determined from the `magnification map' which gives
the total magnification as a function of source position on the source
plane. The magnification map provides information on the number of
images, the calculated magnifications and the lensing cross sections
(cf. M\"oller \& Blain 1998). We compute the magnification map on the
source plane for the simulated groups using ray tracing; the results
for two models are shown in Fig.\,\ref{mapsstrong}.  Comparison of
the two panels in Fig.\,\ref{mapsstrong} shows qualitatively how the
strong lensing properties of a group member depend on the relative
masses of the galaxies within the group. The main
differences/similarities in the magnification maps are:

\begin{enumerate}

\item The area inside the astroid shaped caustic is larger for the model
with a more massive group halo.  Sources that lie inside the astroid
shaped caustics, seen in Fig.\,\ref{mapsstrong}(a) \& (b), are imaged
into four magnified images. Therefore, in this particular
configuration, the lensing galaxy is more likely to produce quadruple
images if it is part of a group with a massive halo than if it is part
of a group without such a halo.

\item The astroid shaped caustic line is longer for a more massive
group halo. The probability that a small background source is
magnified strongly is, to first order, proportional to the length of
the caustic. Extended caustics are therefore more likely to produce
high-magnifications, and so, in this particular configuration, the
lensing galaxy is more likely to produce strongly magnified images if
it is part of a group with a massive halo.

\item The area inside the outer, circular caustic that surrounds the
astroid shaped caustic, is independent of the mass distribution of the
group. Sources that lie outside the area enclosed by this caustic are
not multiply imaged, and, therefore, if observational magnification
bias is ignored, the total strong lensing cross section is not
strongly dependent on the mass distribution of the group.

\end{enumerate}

From this, we conclude that for individual lens systems that have
neighboring groups the details of the mass distribution in the group 
is expected to have a significant effect on the magnifications and 
image geometries.

\subsubsection{Lens modeling and time delays}

Many strong gravitational lens systems have been used to estimate the
Hubble parameter $H_0$ through a measurement of their time delay
(e.g. Koopmans et~al. 2000). Uncertainties in the derived value of
$H_0$ are caused mainly by inaccuracies and uncertainties in models
for the lensing potential. Many of these lens systems are part of, or
lie near, a group (Kundic et~al. 1997b), and it is therefore
important to quantify the effect of the group on the measured time delay. 
Once again ray-tracing routines are used to compute the time delays for
the various configurations.

\begin{figure}
\begin{center}
\psfig{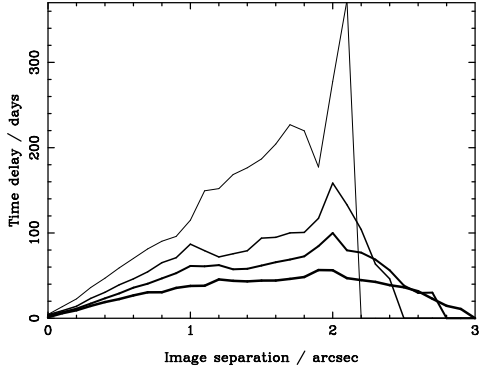}
\caption{The time delay as a function of image separation for the
different group models tabulated in Table\,2. The 
curves are for models A, B, C and D with increasing line thickness. As in
Fig.\,10 the main lensing galaxy has the same properties 
in all cases (galaxy 3, Model C in Table\,2).}
\label{timedel}
\end{center}
\end{figure}

Fig.\,\ref{timedel} shows the time delay as a function of image
separation for the four different group mass distributions listed in
Table\,\ref{table_models}.  Despite the fact that the properties of
the main galaxy are identical in all three panels, there is a
significant variation in the time delay between different group
models. The plot shows that the group potential itself has a great
effect on the time delay; the presence of a massive group halo leads
to smaller maximum image separations and therefore larger time delays.
Since $\Delta T\propto H_0^{-1}$, this has important consequences for
the determination of $H_0$ from such systems. For example, for a lens
system with image separations of $2"$, we estimate that the value of
$H_0$ deduced from a time delay of about 80 days will vary from
$50\,\mathrm{km\,s^{-1}\, Mpc^{-1}}$, for a 70\% halo to
$100\,\mathrm{km\,s^{-1}\,Mpc^{-1}}$ for no group halo. Therefore,
depending on the relative mass of a group halo, the value of $H_0$ may
be seriously underestimated if the group halo is not included in the
lens modeling.

\subsection{Statistical strong lensing}
\label{stronglens_stat}

In the following we investigate the effect of the group mass
distribution on strong lensing statistics using the ray tracing
simulations and a sample of 100 random groups. In each group a single
galaxy at position $\vec{r_{\mathrm{l}}}$ is chosen to be the main
lensing galaxy and we determine the magnifications, time-delays and
image geometries for the lensing galaxy in each group, averaging the
results for the whole sample.  The groups are generated as described
in Sec.~2, except that we set the ellipticities of the individual
galaxies $\epsilon=0$ in the interest of computational speed.  Our
results will not affected by the simplifying assumption that
$\epsilon=0$; since non-zero ellipticities introduce only an
additional statistical error that is proportional to $\sqrt{N}$, where
$N$ is the number of lens systems in the sample.

For each system we obtain statistical information from the image
information stored on a grid in the source plane, as described in
Sec.~3.1.2.  In order to simulate observational selection
effects, we also produce one set of results that only includes images
with magnification ratios smaller than 20 and separations larger than
$0.1"$.  We do not include magnification bias explicitly. This makes the
results discussed below conservative, as magnification bias will
increase the effect of groups on statistical lensing properties, since
highly magnified sources are more probable in lens systems with
substantial external shear.

\subsubsection{Multiplicity of images, image separations and magnification ratios}

We determined the image separations and magnification ratios for all
image pairs for all 100 groups in the simulated sample. Investigating the
statistics of the number of images, we found that changing the mass
distribution inside the group had little effect; the cross sections
for lensing into three, four and five images varied by less than
10\%. This shows that even though image multiplicities of individual
systems may be influenced strongly by the particular direction and
magnitude of the shear, the average effect for a large sample of stacked 
lens systems is small.

The maximum image separations depend on the projected mass contained
inside the smallest circle that contains all images (Schneider et
al. 1992). Therefore, external shear that involves only a contribution
to $\gamma$ will not affect the maximum image separations unless the
image multiplicities are increased. The presence of a group will
affect the maximum image separations only if there is a significant
contribution to the mean $\kappa$ from the group.

\begin{figure}
\begin{center}
\psfig{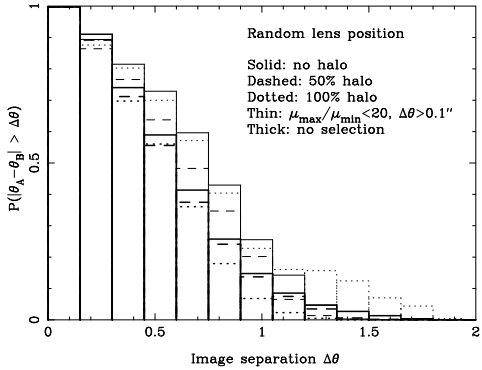}
\caption{The distribution of image separations expected for a large
sample of strong lens systems. The histograms are obtained by binning
the angular separation
$\Delta\theta=|\vec{\theta_{\mathrm{A}}}-\vec{\theta_{\mathrm{B}}}|$
for each pair of images of a given source. Each lensing galaxy is at 
a random position within the group, determined using equation 
\ref{eq_number}. The statistics only include images that are separated 
by more than $0.1"$ and have a magnification ratio less than 1 : 20.}  
\label{avsep}
\end{center}
\end{figure}

Fig.\,\ref{avsep} shows the distribution of image separations for
different group mass distributions for four different choices of
the relative position between lensing galaxy and group. The distance
of the lensing galaxy from the group center is determined randomly
from the distribution given by eq.\,\ref{eq_number}. The figure shows
that, as expected, massive group halos lead to slightly larger
separations. This effect is in principle detectable, given a sample of
100 appropriate lens systems. However, in practice some additional
information is needed to disentangle the degeneracy between a
contribution to $\kappa$ due to a separate group halo component and
due to a more massive lens galaxy.  Fig.\,\ref{avsep} also shows the
significant effect of selection criteria on lensing statistics. If
useful information about the lens population is to be gained from lens
statistics the observational selection criteria need to be understood.

The magnification ratio
$\mu_{\mathrm{r}}=\mu_{\mathrm{A}}/\mu_{\mathrm{B}}$ for two images
$A$ and $B$ is given by eq.\,12. In the case
$\kappa_{\mathrm{A}}\approx\kappa_{\mathrm{B}}\approx0$, the
magnification ratio can be approximated as
$\mu_{\mathrm{r}}=(1-\gamma_{\mathrm{B}}^2)/(1-\gamma_{\mathrm{A}}^2)$. If
$\vec{\gamma}$ varies significantly over distances of the order of the
image separations ($\approx 1$~arcsec), the magnification ratios are
expected to be larger than if $\vec{\gamma}$ varies only
slightly. Thus, any variation of the external shear due to differences
of the mass distribution of the group may change the distribution of
magnification ratios.

\begin{figure}
\begin{center}
\psfig{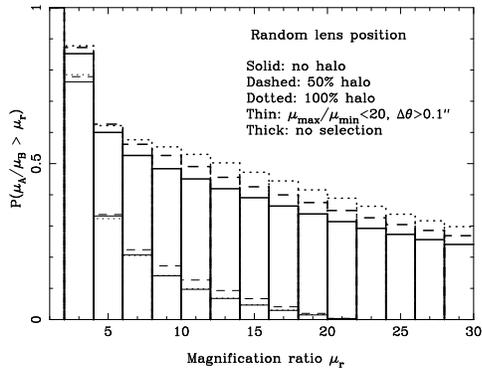}
\caption{The magnification ratio distribution expected for a large sample of strong lens systems. 
The histograms are obtained by binning the magnification ratio for each pair of images of a given source.
The positioning of the lens galaxy is random and the line styles are as in Fig.\,\ref{avsep}.}  
\label{avmag}
\end{center}
\end{figure}

Fig.\,\ref{avmag} shows histograms of the distribution of
magnification ratios for different group mass distributions. The
figure shows that massive group halos lead to slightly larger
magnification ratios, but the effect is small ($<15$\%). As in
Fig.\,\ref{avsep} observational selection effects change the
statistics significantly, even for relatively low magnification ratios
of $\mu_{\mathrm{r}}\approx5$.

\subsubsection{Effect on the time-delay}

Now we assess the effect of the detailed mass distribution on the statistical 
time-delay of a larger sample of lens systems (note that the time-delay
was determined as described in Sec.3.1.2).

\begin{figure*}
\begin{center}
\psfig{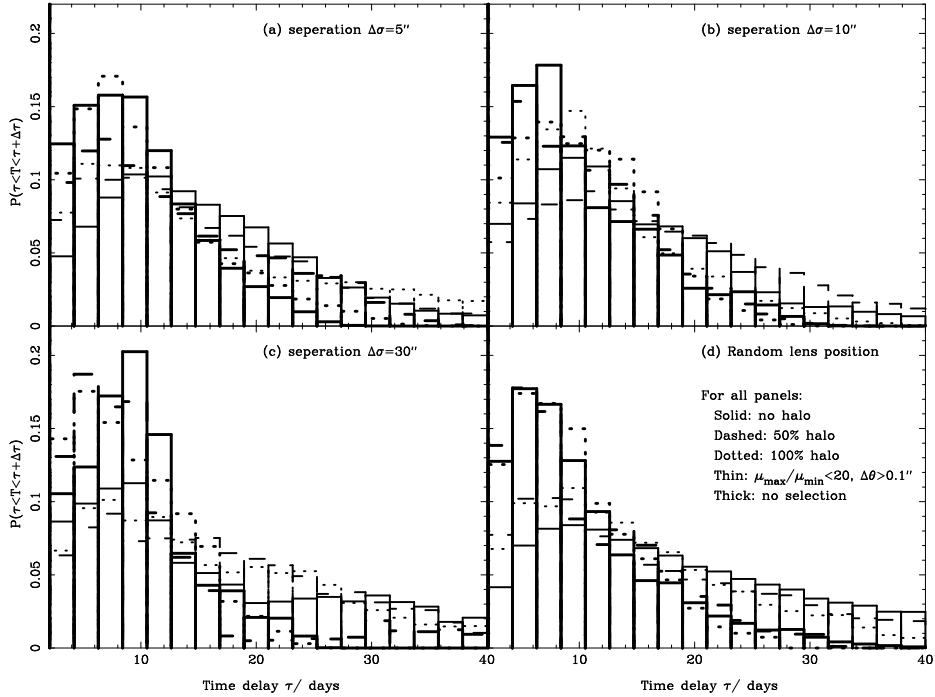}
\caption{The time-delay statistics for different halo masses, from
simulations of 100 group systems. The histograms show the fraction of
systems with a given time-delay that are expected in a large sample of
strong lens systems which are associated with compact groups.  The
panels (a)-(c) show the results for three different separations
between the lensing galaxy and the group center, $\delta\phi$.  In
panel (d) the distance of the lensing galaxy from the group center is
determined randomly from the distribution given by
eq.\,\ref{eq_number}.}
\label{timestat}
\end{center}
\end{figure*}

For each source, the maximum time-delay between image pairs is binned
to create the histograms shown in Fig.\,\ref{timestat}.  Panels
(a)-(c) show the results for separations
$\delta\phi=|\vec{r_{\mathrm{h}}}-\vec{r_{\mathrm{l}}}|$ between the
lensing galaxy and the group center at $\vec{r_{\mathrm{h}}}$ of
$\delta\phi=5"$, $\delta\phi=10"$ and $\delta\phi=30"$ respectively.
In panel (d) the distance of the lensing galaxy from the group center
is determined randomly from the distribution given by
eq.\,\ref{eq_number}.  Strong lens systems in which the main lensing
galaxy is associated with a group that contains a massive group halo
statistically have larger time-delays, whereas lens systems in groups
without a massive halo will have a more strongly peaked distribution
of time-delays with a maximum around the corresponding average
Einstein radius. Thus, measuring the time-delays of a sample of about
100 lens systems associated with compact groups could provide
information on the mass distribution in groups, provided the value of $H_0$
is known.  If the value of $H_0$ is to be deduced from a statistical
sample of lens galaxies, care has to be taken to account fully for the
presence of groups. If a significant number of lens systems used to
determine $H_0$ are near groups with massive halos, and this is not
taken into account, the value of $H_0$ can be underestimated
even by upto factors of 2 or more!.

\section{Conclusions}
\label{groups_discuss}

Even though velocity dispersions can be measured to constrain the
total mass of groups (Mahdavi et~al. 2000), the mass distribution
inside groups is currently not well known. If additional assumptions
about the correlation between X-ray luminosity and mass density are
made, details of the dark matter distribution can be obtained from
sensitive, high-resolution X-ray images. Only gravitational lensing
provides a direct probe of the surface mass density and its spatial
distribution. In this paper, we demonstrate using numerical
techniques, that weak and strong gravitational lensing can be used to
constrain both the total mass and the details of the mass distribution
in groups.

In summary, the results for weak lensing properties groups are as
follows:

\begin{enumerate}

\item The weak lensing shear signal of groups is about 3 per cent, and
varies by up to a factor of 2 for different mass distributions. More
importantly, the ratio of the tangential shear at small radii to that
at larger distances varies by a factor of a few depending on the mass
contained in the group halo.

\item This effect is not detectable for individual groups, but if
about $100$ groups are stacked, then the shear signal around the
individual member galaxies can be determined with sufficient accuracy
to distinguish different mass profiles. The stacking of a large number
of groups also decreases the noise due to cosmic variance, which is of
the same magnitude as the shear signal for individual groups, to a
level of about one per cent of the total group signal.

\item Averaging the shear signal around individual group members has
many practical advantages; the measured shear at larger radii provides
information on the total group mass, whereas the average shear close
to the galaxies measures the galaxy mass fraction.

\item The qualitative results (features in radially averaged
tangential shear profile) are independent of the form of the number
density profile assumed, the halo scale length and possible projection
effects. The level of the signal depends on the details of the
assumed density profile, halo scale length and redshifts; however, the
form of the shear profile as a function of radius remains the same and
is determined only by the halo mass fraction.

\item With new instruments, like ACS (Advanced Camera for Surveys) on
$HST$ or the $NGST$ (Next Generation Space Telescope), it should be
possible to determine the shear to a sufficient accuracy, so that it
will be possible to distinguish different group mass distributions. In
particular, it should be viable to determine whether groups possess
a significant large scale dark halo.

\end{enumerate}
The synopsis of our strong lensing results are:

\begin{enumerate}

\item In the strong lensing regime the presence of groups and the mass
distribution within the group, can affect the magnification maps and caustic
structure significantly for an individual lens in the vicinity.  The
observed time-delay is particularly sensitive to the details of the mass
distribution in the surrounding group. This systematic error needs to be 
taken into account when making estimates of $H_0$ from time-delay measurements.

\item In individual lens systems, the probability of multiple imaging
into 3 or more images may be increased in cases where the main lensing
galaxy is part of or very close to a group. In particular, lens models
which do not take the presence of the group into account are likely to
underestimate the cross section for high-image multiplicities.

\item Statistically, the magnification ratio of images with large
separations is larger for lensing galaxies that are part of groups
with a massive halo.

\item The statistics of time-delays can also be used to constrain the
mass distribution in groups; lens systems in groups with a high halo
mass fraction will on average have larger time-delays.

\end{enumerate}
Weak and strong gravitational lensing studies can provide important
constraints on the mass content and distribution of mass in groups of
galaxies.

\begin{acknowledgments}
OM acknowledges the Lensnet TMR network for support. JPK acknowledges 
support from the CNRS. AWB acknowledges financial support from the 
Raymond and Beverly Sackler foundation as part of the Foundation's 
Deep Sky Initiatve programme at the IoA.
\end{acknowledgments}

\vfill\eject
\end{document}